\documentclass[10pt,aps,pra,twocolumn,english,superscriptaddress]{revtex4-2}
\usepackage{amssymb}
\usepackage{color}
\usepackage{graphicx}
\usepackage{bbold}
\usepackage{bm}
\usepackage{amsmath}
\usepackage{amsfonts}
\usepackage{amssymb}
\usepackage{braket}
\usepackage{upgreek}
\usepackage{xcolor}
\usepackage{fancyhdr}
\usepackage{float}
\usepackage{mathrsfs}
\usepackage{siunitx}
\usepackage{textcmds}
\usepackage{makecell}
\usepackage{CJKutf8}
\usepackage{tikz}

\usepackage[normalem]{ulem}

\definecolor{burgundy}{RGB}{97,0,35}
\definecolor{marine}{RGB}{4,46,96}
\definecolor{greendark}{RGB}{3,53,0}
\usepackage{hyperref}
\hypersetup{%
    colorlinks=true,
    urlcolor=marine,
    linkcolor=marine,
    citecolor=marine
    }
\usepackage{lipsum}

\newcommand{\Sr}[1]{$^{#1}$Sr} 
\newcommand{\RedMotTransition}{${\mathrm{5s^2} ^1\mathrm{S}_0} - {\mathrm{5s5p} ^3\mathrm{P}_1}$}
\newcommand{\BlueMotTransition}{${\mathrm{5s^2} ^1\mathrm{S}_0} - {\mathrm{5s5p} ^1\mathrm{P}_1}$}
\newcommand{\RepumpTransition}{${\mathrm{5s4d} ^1\mathrm{D}_2} - {\mathrm{5s8p} ^1\mathrm{P}_1}$}
\newcommand{\RepumpTransitionSixP}{${\mathrm{5s4d} ^1\mathrm{D}_2} - {\mathrm{5s6p} ^1\mathrm{P}_1}$}
\newcommand{\RepumpRedTransitions}{${\mathrm{5s5p} ^3\mathrm{P}_{0,2}} - {\mathrm{5s6s} ^3\mathrm{S}_1}$}
\newcommand{\StateOneSZero}{${\mathrm{5s^2} ^1\mathrm{S}_0}$}
\newcommand{\StateOneDTwo}{${\mathrm{5s4d} ^1\mathrm{D}_2}$}
\newcommand{\StateFivePOnePOne}{${\mathrm{5s5p} ^1\mathrm{P}_1}$}
\newcommand{\StateSixPOnePOne}{${\mathrm{5s6p} ^1\mathrm{P}_1}$}
\newcommand{\StateEightPOnePOne}{${\mathrm{5s8p} ^1\mathrm{P}_1}$}

\newcommand{\StateThreePOne}{${\mathrm{5s5p} ^3\mathrm{P}_{1}}$}
\newcommand{\StateThreePOneTwo}{${\mathrm{5s5p} ^3\mathrm{P}_{1,2}}$}
\newcommand{\StateThreePZeroTwo}{${\mathrm{5s5p} ^3\mathrm{P}_{0,2}}$}

\newcommand{\BlueMotLinewidth}{$\SI{30}{\mega\hertz}$}
\DeclareSIUnit\gauss{G}

\definecolor{lime}{HTML}{A6CE39}
\DeclareRobustCommand{\orcidicon}{
	\begin{tikzpicture}
	\draw[lime, fill=lime] (0,0) 
	circle [radius=0.16] 
	node[white] {{\fontfamily{qag}\selectfont \tiny ID}};
	\draw[white, fill=white] (-0.0625,0.095) 
	circle [radius=0.007];
	\end{tikzpicture}
	\hspace{-0.3mm}
}

\foreach \x in {A, ..., Z}{\expandafter\xdef\csname orcid\x\endcsname{\noexpand\href{https://orcid.org/\csname orcidauthor\x\endcsname}
			{\noexpand\orcidicon}}
}

\begin{document}

\renewcommand{\figurename}{Fig.}
\renewcommand{\tablename}{Table}

\title{Optical pumping of ${\mathrm{5s4d} ^1\mathrm{D}_2}$ strontium atoms for laser cooling and imaging}

\author{Jens Samland$\orcidC{}$}
\affiliation{Van der Waals-Zeeman Institute, Institute of Physics, University of Amsterdam, Science Park 904, 1098XH Amsterdam, The Netherlands}
\affiliation{Physik-Department, Technical University of Munich, James-Franck-Str.1,
85748 Garching, Germany}
\author{Shayne Bennetts$\orcidE{}$}
\affiliation{Van der Waals-Zeeman Institute, Institute of Physics, University of Amsterdam, Science Park 904, 1098XH Amsterdam, The Netherlands}
\author{Chun-Chia Chen (陳俊嘉)$\orcidA{}$}
\affiliation{Van der Waals-Zeeman Institute, Institute of Physics, University of Amsterdam, Science Park 904, 1098XH Amsterdam, The Netherlands}
\author{Rodrigo Gonz\'{a}lez Escudero$\orcidB{}$}
\affiliation{Van der Waals-Zeeman Institute, Institute of Physics, University of Amsterdam, Science Park 904, 1098XH Amsterdam, The Netherlands}
\author{Florian Schreck$\orcidF{}$}
\affiliation{Van der Waals-Zeeman Institute, Institute of Physics, University of Amsterdam, Science Park 904, 1098XH Amsterdam, The Netherlands}
\affiliation{QuSoft, Science Park 123, 1098XG Amsterdam, The Netherlands}
\author{Benjamin Pasquiou$\orcidD{}$}
\email[]{SrRepump448nm@strontiumBEC.com}
\affiliation{Van der Waals-Zeeman Institute, Institute of Physics, University of Amsterdam, Science Park 904, 1098XH Amsterdam, The Netherlands}
\affiliation{QuSoft, Science Park 123, 1098XG Amsterdam, The Netherlands}
\affiliation{CNRS, UMR 7538, LPL, F-93430, Villetaneuse, France}
\affiliation{Laboratoire de Physique des Lasers, Universit\'e Sorbonne Paris Nord, F-93430, Villetaneuse, France}

\date{\today}

\begin{abstract}
We present a faster repumping scheme for strontium magneto-optical traps operating on the broad \BlueMotTransition~laser cooling transition. Contrary to existing repumping schemes, we directly address lost atoms that spontaneously decayed to the \StateOneDTwo~state, sending them back into the laser cooling cycle by optical pumping on the \RepumpTransition~transition. We thus avoid the $\sim \SI{100}{\micro \second}$-slow decay path from \StateOneDTwo~to the \StateThreePOneTwo~states that is part of other repumping schemes. Using one low-cost external-cavity diode laser emitting at $\SI{448}{nm}$, we show our scheme increases the flux out of a 2D magneto-optical trap by $\SI{60}{\percent}$ compared to without repumping. Furthermore, we perform spectroscopy on the \RepumpTransition~transition and measure its frequency $\nu_{\mathrm{^{88}Sr}} = (668917515.3 \pm 4.0 \pm 25) \, \si{MHz}$. We also measure the frequency shifts between the four stable isotopes of strontium and infer the specific mass and field shift factors, $\delta \nu_\text{SMS} ^{88,86} = \SI{-267(45)}{\mega \hertz}$ and $\delta \nu_\text{FS} ^{88,86} = \SI{2(42)}{\mega \hertz}$. Finally, we measure the hyperfine splitting of the \StateEightPOnePOne~state in fermionic strontium, and deduce the magnetic dipole and electric quadrupole coupling coefficients $A = \SI{-4(5)}{MHz}$ and $B = \SI{5(35)}{MHz}$. Our experimental demonstration shows that this simple and very fast scheme could improve the laser cooling and imaging performance of cold strontium atom devices, such as quantum computers based on strontium atoms in arrays of optical tweezers.
\end{abstract}

\begin{CJK*}{UTF8}{min}
\maketitle
\end{CJK*}

\section{Introduction}
\label{Sec:Introduction}

Laser cooling and trapping techniques such as magneto-optical traps (MOTs) play a pivotal role in cold atom and molecule experiments. As a basic component they enable a wide range of quantum technologies such as optical clocks \cite{Nemitz, campbell_fermi-degenerate_2017}, atom interferometers \cite{dutta_continuous_2016, cheiney_navigation-compatible_2018, menoret_gravity_2018}, quantum simulators and quantum computers, e.g. using arrays of optical tweezers \cite{Saffman2016QuCompRydbProg}. The success of these experiments relies on a combination of the right species with the right laser cooling technique. In recent years, laser-cooled alkaline-earth elements, and in particular strontium, have become widely used platforms for next-generation clocks, interferometers, and quantum computers \cite{Nemitz2016RatioYbSrClock, Abe2021MAGIS, Cooper2018, Norcia2019TweezerClock}. Their success can be attributed to their atomic level structure featuring a broad optical transition perfect for fast laser cooling, and narrow transitions between singlet and triplet states. These provide clock transitions as well as laser-cooling to $\si{\micro \kelvin}$ temperatures and even down to quantum degeneracy \cite{Chen2022CWBEC, Hu2017BECLaserCooling, Stellmer2013LaserCoolingToBEC}. Continued development of new quantum technologies relies on the continued improvement of laser cooling techniques.

In the case of strontium, the broad \BlueMotLinewidth-wide \BlueMotTransition~transition used for the first laser cooling stages is not fully closed. Each cycle through this transition has a probability of one out of 50000 \cite{Hunter1986, Kwela1988Elec1D2to1S0Sr, Xu2003, Jackson2020NumberResolvedTweezer} or one out of 20000 \cite{Werij1992, Porsev2001AtomicTheoSr, Cooper2018} for an atom in the excited \StateFivePOnePOne~state to decay into the \StateOneDTwo~state, leading to losses without an additional repumping scheme. After some time spent in \StateOneDTwo, the atom further decays into the \StateThreePOneTwo~states, from which one typically repumps it back into the \BlueMotTransition~laser cooling cycle. Unfortunately, the lifetime of the \StateOneDTwo~state is rather long, $\SI{300}{\micro \second}$ \cite{Bauschlicher1985Theo1D2decay, Husain1988ExpLifetime1D2, Vogel1999PhDThesis, Bidel2022PhDThesis}. Such a long time spent in an unaddressed state is potentially detrimental when a fast cycling transition is required in order not to lose the atom, e.g. for a Zeeman slower, a 2D MOT, or a fast and reliable single atom detection \cite{Nosske2017Sr2DMOT, Feng2023HighFluxSr}.

Taking the example of fluorescence imaging single strontium atoms trapped in optical tweezers, each atom needs to scatter many photons in order to achieve a high readout fidelity \cite{Covey2019Tweezer2000}. However, an atom that decays into the \StateOneDTwo~state has enough time to escape the tweezers region, if the tweezers is not confining for \StateOneDTwo. Strikingly, several experiments with tweezers have chosen to use light at $515-532 \, \si{\nano \meter}$ \cite{Cooper2018, Jackson2020NumberResolvedTweezer}, because of the available laser power, the small wavelength and since $515 \, \si{\nano \meter}$ allows to create traps with magic wavelength for the \RedMotTransition~transition \cite{Cooper2018, Norcia2018SrTweezer}. Moreover, the polarizability at such short wavelength can be suitable for trapping strontium atoms excited to Rydberg states \cite{Mukherjee2011ManyBodyRydbergAE, Madjarov2021PhDThesis, Wilson2022TrapRydbergAE}. However, tweezers at these wavelengths create a repulsive potential for the \StateOneDTwo~state that expels the atoms. Even when the tweezers wavelength is chosen to be confining, the decay toward the \StateOneDTwo~state can remain an inconvenience as one needs to wait for the atoms to decay to the \StateThreePOneTwo~states.
For example, fast single atom fluorescence detection, which is useful for quantum error correction, error erasure conversion \cite{Ma2023MidCircuitErasure, Scholl2023ErasureConversion}, or the spatially resolved imaging of non-trapped atoms \cite{Buecker2009PropagImaging, Holten2022Cooper2DFermi}, becomes less reliable as a result of this wait time. Finally, as atoms decay from the \StateOneDTwo~state, one may also need to use several repumping lasers to plug every decay path \cite{Dinneen1999, Zhang2020Repump2.6mum}. Such a solution is technologically demanding and thus can affect the robustness of a quantum device using ultracold strontium. To overcome some of these limitations, a few experiments have started investigating the use of the \RedMotTransition~transition for high-fidelity imaging \cite{Urech2022NarrowLineTweezer}.

In this article, we tackle the challenge of the loss channel to the \StateOneDTwo~state with a fast and simple repumping scheme to transfer atoms directly from \StateOneDTwo~back into the ground state using the intermediate \StateEightPOnePOne~state. For single atom detection, this scheme removes the concern for the confining character of the tweezer light for the states involved and the concern about reliably high scattering rate, as atoms do not have enough time to leave the trap before our fast repumping. Excitingly, this scheme also opens the opportunity for shelving atoms in the clock or qubit \StateThreePZeroTwo~states, while imaging ground state atoms without contamination of the sample through the decay path via the \StateOneDTwo~state \cite{Okuno2022TweezerYb1S03P2Qbit, Pagano2022TweezerRydbergSr, Trautmann2023SpectroSr1S03P2}. According to known atomic data \cite{Werij1992, Sansonetti2010}, our scheme should be able to repump more than $\SI{96}{\percent}$ of the atoms using decay paths faster than $\SI{100}{\nano \second}$. Let us note that the remaining atoms will follow slightly slower decay paths, while a fraction of them could fall into long-lived triplet states, and the branching ratios toward these states still remain to be measured. The suitability of our scheme will thus strongly depend on the intended application. Pragmatically, our scheme can be realised with a single, low-cost, external-cavity diode laser at $\SI{448}{nm}$, which relaxes the technological requirements for ultracold Sr devices. By implementing this scheme, we show that the atom flux out of a 2D MOT operating on the \BlueMotTransition~transition is increased by about $\SI{60}{\percent}$ compared to without repumping. 

This article is structured as follows: in Section~\ref{Sec:StateOfTheArt}, we present the state of the art by listing the various repumping schemes that have been experimentally demonstrated in recent years, and we propose our novel scheme. Then in Sec.~\ref{Sec:ExperimentalProtocol}, we describe our experimental setup. In Sec.~\ref{Sec:Spectroscopy}, we present spectroscopic data for the \StateOneDTwo~- ~\StateEightPOnePOne~transition. This includes the determination of the absolute frequency of the transition, the isotope shifts between the four stable isotopes of strontium, and the hyperfine structure present for the fermionic isotope $^{87}$Sr. Finally, in Sec.~\ref{Sec:Repumping} we characterize the repumping efficiency of our scheme applied to a 2D MOT.

\section{State of the art and our solution}
\label{Sec:StateOfTheArt}

\begin{figure}[tb]
	\centering
	\includegraphics[width=0.98\columnwidth]{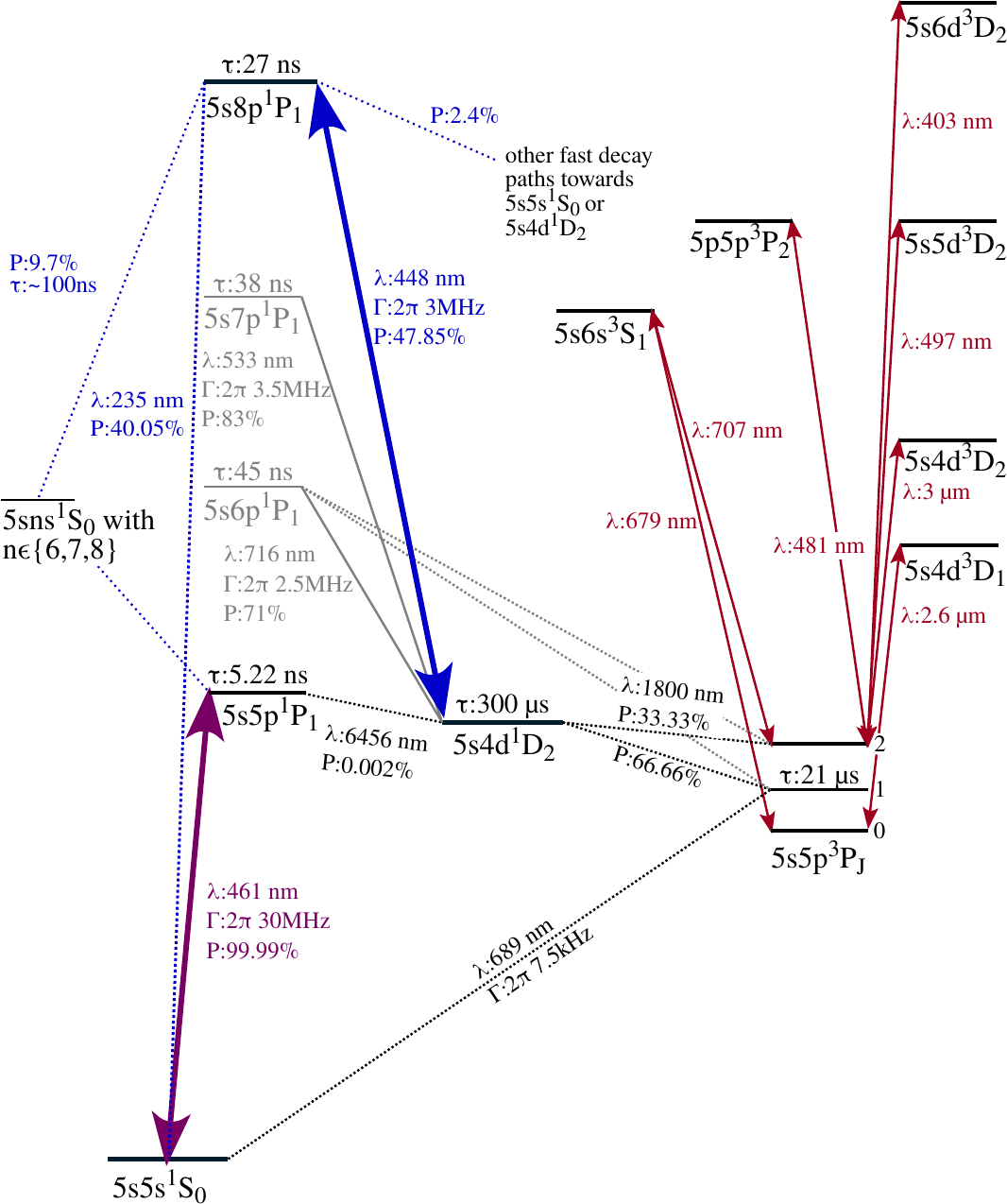}
	\caption{Low-lying strontium levels and transitions. Typically, atoms are cooled using the \BlueMotTransition~transition, with a one out of 50000 \cite{Hunter1986, Kwela1988Elec1D2to1S0Sr, Xu2003, Jackson2020NumberResolvedTweezer} or one out of 20000 \cite{Werij1992, Porsev2001AtomicTheoSr, Cooper2018} chance for atoms to decay from the \StateFivePOnePOne~state into the \StateOneDTwo~state. Our scheme optically pumps these lost atoms to the \StateEightPOnePOne~state, from which they can quickly decay back into the cooling cycle. The diagram also shows states that are connected to the 5s5p$^3$P$_{\textrm{J}}$ states and that are addressed in other commonly used repumping schemes. We took the transition data from \cite{Poli2005, Werij1992} and indicate transition probabilities, decay rates and lifetimes when relevant for further understanding.}
	\label{Fig:ReducedTermDiagram}
\end{figure}

Here we first describe existing repumping schemes, then present ours and motivate this choice. Figure~\ref{Fig:ReducedTermDiagram} shows the broad, \BlueMotLinewidth-wide transition at $\SI{461}{nm}$ between \BlueMotTransition, which is typically used for laser cooling strontium. However, while cycling on this transition, an atom has a one out of 50000 \cite{Hunter1986, Kwela1988Elec1D2to1S0Sr, Xu2003, Jackson2020NumberResolvedTweezer} or one out of 20000 \cite{Werij1992, Porsev2001AtomicTheoSr, Cooper2018} chance of decaying from the upper \StateFivePOnePOne~state into the \StateOneDTwo~state, thus leaving the cooling/imaging cycle. In order to bring these atoms back into the \BlueMotTransition~cycle, many repumping schemes have been proposed and demonstrated experimentally (see also references in \cite{Hu2019analyzing}):
\begin{itemize}
    \item 5s5p$^3$P$_{0,2}$ $\to$ 5s6s$^3$S$_1$ at $\SI{679}{nm}$ and $\SI{707}{nm}$ in \cite{Dinneen1999}
    \item 5s5p$^3$P$_2$     $\to$ 5s4d$^3$D$_2$ at $\SI{3011}{nm}$ in \cite{Mickelson2009RepumpingAS}
    \item 5s5p$^3$P$_2$     $\to$ 5s5d$^3$D$_2$ at $\SI{497}{nm}$ in \cite{Stellmer2014ReservSpectro, Schkolnik2019extended, Moriya2018, sorrentino2006laser}
    \item 5s5p$^3$P$_2$     $\to$ 5s6d$^3$D$_2$ at $\SI{403}{nm}$ in \cite{Stellmer2014ReservSpectro, Moriya2018}
    \item 5s5p$^3$P$_2$     $\to$ 5p$^2$\,$^3$P$_2$ at $\SI{481}{nm}$ in \cite{Hu2019analyzing}
    \item 5s5p$^3$P$_{2}$ $\to$ 5s6s$^3$S$_1$ at $\SI{707}{nm}$ and 5s5p$^3$P$_{0}$ $\to$ 5s4d$^3$D$_1$ at $\SI{2.6}{\micro \meter}$ in \cite{Zhang2020Repump2.6mum}.
    \label{List:RepumpingSchemes}
\end{itemize}

Many of these schemes are implemented using a single laser source that can be built cheaply using available laser diodes (e.g. \cite{Hu2019analyzing, Moriya2018, Stellmer2014ReservSpectro}). They greatly increase the lifetime of the atoms in the MOT and thus also their numbers. In general, the practical efficiency of a scheme depends on whether the branching ratios between various decay paths are such that most of the atoms can fall back into the laser cooling cycle by addressing only the fewest number of optical transitions~\cite{Xu2003}. Even for the fermionic isotope \Sr{87}, whose non-zero nuclear spin $I = 9/2$ generates a complex hyperfine structure with many more transitions, these schemes can still be advantageous, as long as the hyperfine structures of the states at play are not too widely spaced. In this case, it is not too technologically involved to address all relevant hyperfine transitions, using a single laser source~\cite{Mickelson2009RepumpingAS, Stellmer2014ReservSpectro}. However, all of these schemes rely on the optical pumping of atoms from the $\mathrm{5s5p} ^3\mathrm{P}_{\textrm{J}}$ state manifold to higher lying intermediate states, before the atoms decay to the ground state via the \RedMotTransition~transition, see Fig.~\ref{Fig:ReducedTermDiagram}.

Accordingly, they require that atoms decay from the \StateOneDTwo~state to the \StateThreePOneTwo~states. The characteristic time of this decay is $\SI{303(10)}{\micro \second}$ \cite{Vogel1999PhDThesis, Bauschlicher1985Theo1D2decay, Husain1988ExpLifetime1D2, Bidel2022PhDThesis}, severely limiting the repumping timescales. By contrast, our scheme uses the \RepumpTransition~transition at $\SI{448}{\nano \meter}$ to directly address atoms in the \StateOneDTwo~state, similar to what has already been demonstrated in Ca \cite{Mills2017RepumpCa}. This way, atoms are swiftly repumped to the ground state, thanks to several decay paths from \StateEightPOnePOne~with characteristic times of about $\SI{100}{\nano \second}$ \cite{Sansonetti2010, Werij1992}, see Fig.~\ref{Fig:ReducedTermDiagram} and Tab.~\ref{Tab:BranchingRatios}. Let us note that some experiments do not continuously shine repumping beams to prevent losses in the MOT. Instead, they make use of the long lifetime of the $\mathrm{5s5p} ^3\mathrm{P}_{2,0}$ states and accumulate atoms in a low-loss reservoir, before applying a repumping pulse that brings atoms toward the \StateOneSZero~ground state \cite{Nagel2003MagneticTrappingSr3P2}. Our repumping scheme is not aimed at these protocols, as it addresses directly the \StateOneDTwo~state.

\begin{table}[H]
	\centering
	\begin{tabular}{l|c|c}
    Decay path from ${\mathrm{5s8p} ^1\mathrm{P}_1}$ & Probability & $1/e$ time\\
     & (in $\si{\percent}$) & (in $\si{\nano \second}$)\\
	\hline
    $\rightarrow {\mathrm{5s^2} ^1\mathrm{S}_0}$  & $40.0$  & $27$ \\
    $\rightarrow {\mathrm{5s4d} ^1\mathrm{D}_2}$  & $47.8$  & $27$ \\
    $\rightarrow {\mathrm{5s6s} ^1\mathrm{S}_0} \rightarrow {\mathrm{5s5p} ^1\mathrm{P}_1} \rightarrow {\mathrm{5s^2} ^1\mathrm{S}_0}$  & $7.1$  & $86$ \\
	$\rightarrow {\mathrm{5s7s} ^1\mathrm{S}_0} \rightarrow {\mathrm{5s5p} ^1\mathrm{P}_1} \rightarrow {\mathrm{5s^2} ^1\mathrm{S}_0}$   & $1.8$  & $51$ \\
    $\rightarrow {\mathrm{5s8s} ^1\mathrm{S}_0} \rightarrow {\mathrm{5s5p} ^1\mathrm{P}_1} \rightarrow {\mathrm{5s^2} ^1\mathrm{S}_0}$   & $0.8$  & $115$ \\
	\end{tabular}
	\caption{Most likely ($ \gtrsim 1\% $) decay paths after excitation to the \StateEightPOnePOne~state using the \RepumpTransition~transition. The second column gives the overall probability for an atom to undergo each path, using the branching ratios from Ref.~\cite{Werij1992} for the singlet states. The branching ratios toward triplets states are not well known, see e.g. Refs.~\cite{Kurosu1990LaserCoolCaSr, Vogel1999PhDThesis, Bidel2022PhDThesis}. The last column gives the sum of the $1/e$ lifetimes of all excited states populated along a decay path. See Ref.~\cite{Werij1992} for other, less likely decay paths, all requiring a few hundreds of nanoseconds.}
	\label{Tab:BranchingRatios}
\end{table}

To address directly atoms in the \StateOneDTwo~state, the use of the ladder of ${\mathrm{5s}n\mathrm{p} ^1\mathrm{P}_1}$ electronic states is a natural choice. The lower lying states on this ladder should ensure a high probability for the atoms to decay to the ground state without getting lost in a forest of lower-lying states, some of which are potentially long lived. In that regard, the state ${\mathrm{5s6p} ^1\mathrm{P}_1}$ has allowed decay paths only toward \StateOneSZero, \StateOneDTwo, and ${\mathrm{5s6s} ^1\mathrm{S}_0}$. Moreover, the ${\mathrm{5s4d} ^1\mathrm{D}_2} - {\mathrm{5s6p} ^1\mathrm{P}_1}$ transition at $\SI{716}{\nano \meter}$ \cite{Grundevik1983} is somewhat directly accessible with today's laser diode technologies. However, this repumping transition has been studied experimentally before, but was found to be rather ineffective at sending atoms back into the cooling cycle \cite{Kurosu1990LaserCoolCaSr, Vogel1999PhDThesis, Bidel2022PhDThesis}. These demonstrations hypothesize significant losses via decays from ${\mathrm{5s6p} ^1\mathrm{P}_1}$ to the triplet \StateThreePOneTwo~states, on top of the decays towards ${\mathrm{5s6s} ^3\mathrm{S}_1}$ and ${\mathrm{5s4d} ^3\mathrm{D}_1}$ from Refs.~\cite{Werij1992, Sansonetti2010}. Nonetheless, Ref.~\cite{Bidel2022PhDThesis} confirms that the overall restoring force from laser cooling on \BlueMotTransition~can be increased by reducing the time atoms spend in \StateOneDTwo. Finally, the branching ratios \cite{Werij1992} for the decay from this state favor a decay back to \StateOneDTwo, meaning that on average several pumping photons will be required before reaching \StateOneSZero. The next state on the ladder, ${\mathrm{5s7p} ^1\mathrm{P}_1}$, has a transition from \StateOneDTwo~at $\SI{533}{\nano \meter}$ suffering from the same branching ratio issue, and possesses additional states reachable via spontaneous decay. Moreover, tunable laser diodes emitting directly at this wavelength are not common. This leaves us with the \StateEightPOnePOne~state as the lowest energy state with good branching ratios toward \StateOneDTwo, see Fig.~\ref{Fig:ReducedTermDiagram} and Ref.~\cite{Werij1992}. Despite being numerous, the decay paths from \StateEightPOnePOne~through singlet states all lead in short times (a few $\SI{100}{\nano \second}$) to the states \StateOneSZero, \StateFivePOnePOne, or \StateOneDTwo~\cite{Werij1992}. Finally, at this wavelength, cheap laser diodes are available. These arguments lead us to investigate the \RepumpTransition~transition as a potential repumping transition.

\section{Experimental setup}
\label{Sec:ExperimentalProtocol}

\begin{figure}[tb]
	\centering
	\includegraphics[width=0.8\columnwidth]{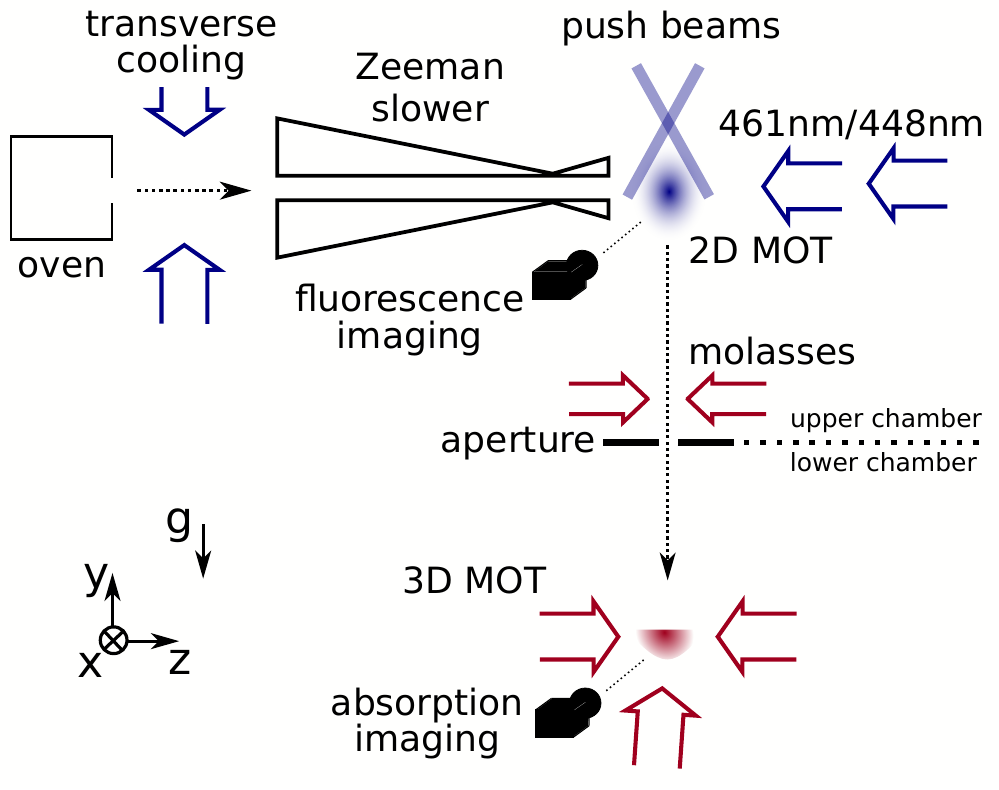}
	\caption{Experimental setup. Atoms from a strontium oven are slowed down by a Zeeman slower, then cooled in a two-dimensional magneto-optical trap employing the cycling \BlueMotTransition~transition at $\SI{461}{nm}$. We use fluorescence imaging to measure the atom number in the MOT. From this 2D MOT, atoms fall down under gravity into a three-dimensional MOT working on the narrow \RedMotTransition~transition at $\SI{689}{nm}$, where they are imaged via absorption imaging. Our repumping laser at $\SI{448}{nm}$ takes the same path as the $\SI{461}{nm}$ light for the Zeeman slower and shines onto the 2D MOT.} 
	\label{Fig:setup}
\end{figure}

Our experimental setup, see Fig.~\ref{Fig:setup}, has been described in detail in \cite{Bennetts2017steady}. We recall here the most important points. The atom source is an oven in which metallic strontium with natural isotopic abundance is heated to about $\SI{470}{\celsius}$. An atomic beam then effuses from a microtube nozzle, is collimated transversely by an optical molasses, and Zeeman slowed to about $\SI{20}{\meter \per \second}$. Both cooling stages use the \BlueMotTransition~transition. The slowed atoms enter a vacuum chamber labeled ``upper chamber", and reach a 2D-MOT whose axis is located $\SI{5}{cm}$ after the slower's exit. This MOT uses the \BlueMotTransition~transition as well and radially cools the atomic sample to a Gaussian velocity spread of about $\SI{0.5}{\meter \per \second}$. The radial magnetic field gradient is $\SI{10}{\gauss \per \centi \meter}$. Atoms can escape the MOT along the non-confining, vertical, $y$ axis.

For characterizing the effect of our repumping scheme on this 2D MOT, we use two techniques. First, in Sec.~\ref{Sec:Spectroscopy} we observe the variations of the MOT fluorescence by imaging it onto a camera located in the same horizontal plane. Second, in Sec.~\ref{Sec:Repumping} we record the atoms that exit the 2D MOT downward, pass into a different vacuum chamber, labeled ``lower chamber", and get captured in a 3D MOT operated on the \RedMotTransition~transition, see Fig.~\ref{Fig:setup} and Ref.~\cite{Bennetts2017steady}. In the present work, we applied the latter technique only to bosonic species. Let us note that we recently applied this method to produce a similar continuous 3D MOT for fermionic strontium \cite{Escudero2021}. When transferring atoms into the lower chamber, we prevent atoms in the 2D MOT from going upward by applying downward propagating ``push" beams, in the presence of which we measure the falling atoms average velocity of $\SI{3}{\meter \per \second}$ \cite{Bennetts2017steady}. These push beams, addressing the \BlueMotTransition~transition, are not present when we perform spectroscopy looking at the MOT fluorescence (Sec.~\ref{Sec:Spectroscopy}), but they are active when we measure the MOT's outgoing flux (Sec.~\ref{Sec:Repumping}). 

As repumping laser, we use a homemade~\cite{Bennetts2019PhDThesis} external cavity diode laser emitting directly at $\SI{448}{nm}$. We choose the \textit{GH04580A2G} laser diode from Sharp, with a \textit{GH13-24V} grating from Thorlabs placed in a Littrow configuration. We stabilize the laser frequency by locking it via the Pound-Drever-Hall technique to a stable optical cavity with finesse $\mathscr{F} \approx 3500$ and a free spectral range of $\SI{1.5}{\giga \hertz}$. The locked-laser frequency drifts by only $\sim \SI{500}{kHz}$ over $\SI{90}{\minute}$, which is suitable for the \RepumpTransition~transition whose linewidth is $\Gamma = 2 \pi \times \SI{3.0(7)}{MHz}$ \cite{Sansonetti2010}. We use a piezoelectric transducer to tune the cavity length and thus the frequency of the locked laser. The cavity resonance can be tuned by up to $\sim \SI{30}{\giga \hertz}$, but the continuous scanning range is limited to the laser's mode-hope-free range of $\sim \SI{3}{\giga \hertz}$.
The laser delivers up to $\SI{4.5}{\milli \watt}$ at $\SI{448}{nm}$. In order to increase the available power, we ``injection lock" a secondary laser diode (same model) using as low as $\SI{150}{\micro \watt}$ from the external cavity diode laser. After coupling into an optical fiber with $\SI{50}{\percent}$ efficiency, we obtain $\SI{16}{\milli \watt}$ to be sent onto the atoms. For simplicity, we combine the repumping beam with the $\SI{461}{nm}$ Zeeman slower light using a polarizing beamsplitter. The repumper polarisation is thus set mostly circular and opposite to the Zeeman slower one. We shape the beam to a $1/e^2$ diameter of $\SI{7.5}{mm}$ in order to cover the 2D MOT cloud, in particular along the non-confining axis.

\section{Spectroscopy}
\label{Sec:Spectroscopy}

We perform spectroscopy of the \RepumpTransition~transition at $\SI{448}{nm}$ using fluorescence imaging of the atoms in the 2D MOT, see Fig.~\ref{Fig:setup}. By scanning the repump laser frequency across resonance, we optically pump atoms from the dark \StateOneDTwo~state back into the cooling cycle, such that these lost atoms contribute again to the fluorescence signal, see Fig.~\ref{Fig:ReducedTermDiagram}. More precisely, having a 2D MOT continuously loaded with atoms of a single, selected isotope, we switch the $\SI{448}{nm}$ repump laser on and scan its frequency over a range of about $\SI{200}{MHz}$ around the \RepumpTransition~resonance. We apply a symmetric triangular voltage ramp on the cavity piezoelectric transducer to scan the laser frequency. Each linear segment of the ramp lasts $\sim \SI{10}{\second}$, chosen slow enough so as to let the population reach a steady state at all times. During the ramp, the laser frequency is recorded by a \textit{WS7-30} wavemeter from HighFinesse, and we scan in total five triangular ramps for averaging. We record the fluorescence signal with a \textit{Blackfly 2.3MP Mono GigE} camera (BFLY-PGE-23S6M-C). We choose its region of interest so as to detect the fluorescence coming only from the upper MOT region, excluding the MOT center. We thus avoid recording an undesired signal from fast atoms traveling along the Zeeman slower's beam but faster than the 2D MOT capture velocity, which would add Doppler broadening to the spectroscopy signal.

\subsection{Absolute transition frequency}
\label{SubSec:Absolutetransitionfrequency}

\begin{figure}[tb]
	\centering
	\includegraphics[width=0.98\columnwidth]{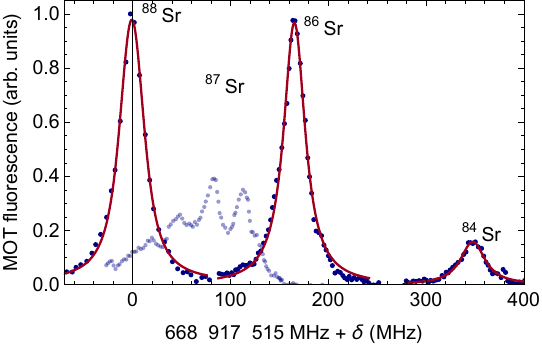}
	\caption{Example spectra of the \RepumpTransition~transition for the four stable isotopes of strontium, via fluorescence of the two-dimensional MOT. In contrast to the three bosonic isotopes \Sr{88}, \Sr{86} and \Sr{84}, the fermionic isotope \Sr{87} has a nuclear momentum of $I=9/2$ and thus exhibits a hyperfine structure. Data (blue points) are fitted with a Voigt profile (red curves) to extract the center frequency and isotope shifts. Note that the experimental parameters are adapted for each isotope in order to improve the signal-to-noise ratio, which means that the fluorescence intensities are not representative of strontium's natural abundance. The background fluorescence of the MOT is offset to zero.} 
	\label{Fig:Spectroscopy}
\end{figure}

We spectroscopically determine the absolute transition frequency of the \RepumpTransition~transition for the bosonic, most abundant isotope \Sr{88}, see Fig.~\ref{Fig:Spectroscopy}. We measure it to be $\nu_{\text{88Sr}} = (668917515.3 \pm 4.0 \pm 25) \, \si{MHz}$, where the first error bar is the statistical uncertainty and the second comes from the accuracy of the wavemeter, see below. NIST provides for this transition the frequency value of $\nu_{\text{NIST}} = \SI{668916207(300)}{MHz}$ \cite{Sansonetti2010}, which is a disagreement with our measurement of more than $\SI{1}{GHz}$. We ensure the accuracy of our measurement by calibrating the wavemeter across the visible spectrum \cite{Samland2019MastersThesis}, similar to the calibration in Ref.~\cite{Stellmer2014ReservSpectro}. We use well-known optical transitions in different elements available in the cold atoms and ions experiments in our laboratory. Namely, we use the $\SI{441}{nm}$ transition in Yb$^+$ \cite{Roberts1999}, the $\SI{461}{nm}$, $\SI{689}{nm}$, and $\SI{497}{nm}$ transitions in Sr \cite{Stellmer2014ReservSpectro}, and the $\SI{780}{nm}$ D2-transition in Rb \cite{Ye1996}. For all these transitions, the wavemeter reflects the literature values within an accuracy of $\SI{25}{MHz}$, which is compatible with the $\SI{30}{MHz}$ accuracy specified by the manufacturer. We set the wavemeter calibration with respect to the \RedMotTransition~transition of strontium at $\SI{689}{\nano \meter}$, for which our reference laser's accuracy is better than $\SI{50}{\kilo \hertz}$ \cite{Courtillot2005SpectroSr}.

In order to determine the transition's frequency from the recorded fluorescence signal, we fit a single Voigt profile to the data, see Fig.~\ref{Fig:Spectroscopy}. We use this profile to take into account the various broadening mechanisms. The MOT finite radial velocity spread of $\SI{0.5}{\meter \per \second}$ \cite{Bennetts2017steady} should contribute to a broadening with a Gaussian width of $\SI{1.1}{\mega \hertz}$. The MOT diameter in the horizontal plane is as big as $\SI{4}{\milli \meter}$, which leads to a Zeeman broadening of $\sim |6| \, \si{\mega \hertz}$ in the magnetic field of the  MOT. The laser beam intensity is $16.5$ times the transition's saturation intensity of $\SI{4.4}{\milli \watt \per \square \centi \meter}$, and thus broadens the transition width from $2 \pi \times \SI{3.0(7)}{\mega \hertz}$ to $2 \pi \times \SI{12.6}{\mega \hertz}$. Combined, these broadening mechanisms are roughly accounting for the fitted Voigt profile, which exhibits a full-width half-maximum of $\sim 2 \pi \times  \SI{30}{\mega \hertz}$, see Tab.~\ref{Tab:ExpectedMeasuredIsotopShifts}.

\begin{table}[tb]
	\centering
	\begin{tabular}{l|l|l|l|l}
    Isotope & \makecell{$\nu_{\text{res}}$ \\ (MHz)} & \makecell{$\delta \nu ^{A,88}$ \\ (MHz)} & \makecell{FWHM \\ (MHz)} \\
	\hline
 	$84$ & $668 917 860.1 \pm 2.6  \pm 25$  & $-344.8 \pm 6.6$  & $29 \pm 5$ \\
  	$86$ & $668 917 682.3 \pm 2.6  \pm 25$  & $-167.0 \pm 6.6$  & $29 \pm 3$ \\
    $87$ & $668 917 598.8 \pm 12.4 \pm 25$  & $-83.5  \pm 16.4$ &            \\
	$88$ & $668 917 515.3 \pm 4.0  \pm 25$  & 0                 & $30 \pm 2$ \\
	\end{tabular}
	\caption{Absolute transition frequencies, isotope shifts, and linewidths from the spectroscopy of the \RepumpTransition~transition for the four stable isotopes of strontium. The given widths are the full-width half-maximum of the fitted Voigt profiles. The absolute frequency for \Sr{87} and the isotope shift between \Sr{88} and \Sr{87} and their uncertainties come from the analysis of Sec.~\ref{SubSec:HyperFineStructure}, and represent the virtual \Sr{87} transition without the hyperfine splitting. The second error bars on the absolute frequencies comes from the specified accuracy of the wavemeter, see Sec.~\ref{SubSec:Absolutetransitionfrequency}. All remaining error bars represent the standard deviation of the fitted parameters from 5 consecutive triangular ramps.}
	\label{Tab:ExpectedMeasuredIsotopShifts}
\end{table}

\subsection{Isotope shifts}
\label{SubSec:IsotopeShifts}

In this section, we spectroscopically determine the isotope shifts of the \RepumpTransition~transition frequency between \Sr{88} and the three other stable isotopes \Sr{87}, \Sr{86}, and \Sr{84}. In order to determine these shifts, we load a single isotope in our 2D MOT by setting the detunings of the Zeeman slower and MOT laser beams to match the well-known isotope shifts of the $\SI{461}{nm}$ cooling transition, see Fig.~\ref{Fig:setup}. We then perform spectroscopy on this isotope by scanning the $\SI{448}{nm}$ laser frequency as in the case of \Sr{88}, see Sec.~\ref{SubSec:Absolutetransitionfrequency}. Subsequently, we proceed to the next isotope by setting the proper laser cooling parameters. In our setup, it takes about five minutes to pass from laser cooling one isotope to another, due to the need for small optimizations such as optical fiber injections and adjustments of MOT parameters. Since the measurement of isotope shifts is relative in nature, and since we keep calibrating regularly our wavemeter, the uncertainty on the isotope shifts is greatly reduced, and is now dominated by the short-term variations from scan to scan. The cooling laser at $\SI{461}{nm}$ is locked by saturated absorption spectroscopy to better than $\SI{5}{\mega \hertz}$, and thus does not drift over long times.

We repeat the same protocol described above for the two bosonic isotopes \Sr{86} and \Sr{84}. The measured values are reported in Tab.~\ref{Tab:ExpectedMeasuredIsotopShifts}. Contrary the other stable isotopes, \Sr{87} is fermionic and exhibits a hyperfine structure, see Fig.~\ref{Fig:Spectroscopy}. This prevents us from fitting a single Voigt profile to the fluorescence signal and determining the isotope shift this way. Instead, we provide the value of $\Delta \nu$ of the virtual transition without hyperfine splitting, obtained through the knowledge of the magnetic dipole and electric quadrupole coupling coefficients A and B, thanks to the analysis of Sec.~\ref{SubSec:HyperFineStructure}.

\begin{figure}[tb]
	\centering
    \includegraphics[width=0.98\columnwidth]{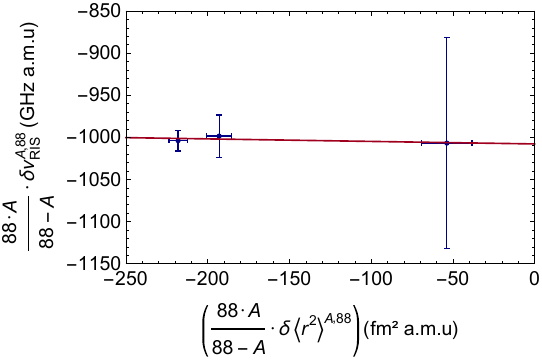}
	\caption{Determination by linear regression of the specific mass shift factor $M_{\text{SMS}}$ and field shift factor $F$ for the \RepumpTransition~transition. A similar analysis was performed for calcium in \cite{Dammalapati2010IsotopeShiftCa}. Values for $\delta \langle r^2 \rangle ^{A,A'}$ are taken from \cite{Angeli2013NuclearTableUpdate}.} 
	\label{Fig:isotopeShiftsLinRegression}
\end{figure}

From the measured isotope shifts $\delta \nu ^{A,A'} = \delta \nu ^{A'} - \delta \nu ^{A}$, we deduce the mass shift $\delta \nu_\text{MS} ^{A,A'}$ and field shift $\delta \nu_\text{FS} ^{A,A'}$ following the analysis for calcium in Ref.~\cite{Mortensen2004Isotope}:
\begin{equation}
    \delta \nu ^{A,A'} = \delta \nu_\text{MS} ^{A,A'} + \delta \nu_\text{FS} ^{A,A'} = M \cdot \frac{A'-A}{A'A} + F \cdot \delta \langle r^2 \rangle ^{A,A'}.
    \label{Eq:isotope_shift}
\end{equation}
Here, $M$ and $F$ are the mass and field shift factors, $A$ and $A'$ are the atomic masses of two isotopes, and $\delta \langle r^2 \rangle ^{A,A'}$ is the nuclear radii change between these two isotopes \cite{Heilig1974ChargeRadii}. The mass shift factor is written as the sum of the normal and the specific mass shift factors $M = M_{\text{NMS}} + M_{\text{SMS}}$, where $M_{\text{NMS}} = \nu \, m_e / m_a$ with $\nu$ the transition frequency, $m_e$ the electron mass and $m_a$ the atomic mass unit. Usually, the residual (or modified) isotope shift $\delta \nu ^{A,A'} _\text{RIS}$ is obtained by subtracting the normal mass shift from the measured isotope shift $\delta \nu ^{A,A'}$. Thus, we can rewrite Eq.~(\ref{Eq:isotope_shift}) as
\begin{equation}
\begin{split}
    \frac{A'A}{A'-A} \delta \nu ^{A,A'} _\text{RIS} &= \frac{A'A}{A'-A}  \delta \nu ^{A,A'} - M_\text{NMS} \\
    &= M_\text{SMS} + F\cdot \frac{A'A}{A'-A} \cdot \delta \langle r^2 \rangle ^{A,A'}
    \label{Eq:modified_isotope_shift}
\end{split}
\end{equation}
and obtain the two factors $M_{\text{SMS}}$ and $F$ as the offset and slope of a linear regression, see Fig.~\ref{Fig:isotopeShiftsLinRegression}. This procedure yields $M_\text{SMS} = \SI{-1008(170)}{\giga \hertz \, \textrm{a.m.u.}}$ and $F = \SI{-0.03(0.83)}{\giga \hertz \per \square {fm}}$, where $\textrm{a.m.u.}$ stands for atomic mass unit, and values for $\delta \langle r^2 \rangle ^{A,A'}$ are taken from Ref.~\cite{Angeli2013NuclearTableUpdate}. To our knowledge, the only experimental determination of an isotope shift of the \RepumpTransition~transition has been made in Ref.~\cite{Lorenzen1982isotopeShiftSrGalvano}. There, the authors use optogalvanic spectroscopy to determine the isotope shift between \Sr{88} and \Sr{86}, $\delta \nu ^{88,86} = \SI{-163(6)}{\mega \hertz}$. This is compatible with our value of $\delta \nu ^{88,86} = -167(6.6) \, \si{\mega \hertz}$. Thanks to our measurement of the other stable isotopes, in addition to confirming the results of Ref.~\cite{Lorenzen1982isotopeShiftSrGalvano}, we are able to provide the specific mass shift factor and the field shift factor. We can further compare our results with the theoretical calculations of Ref.~\cite{Aspect1991IsotopeShiftSr}, which predict $\delta \nu_\text{SMS} ^{88,86} = \SI{-211}{\mega \hertz}$ and $\delta \nu_\text{FS} ^{88,86} = \SI{-37}{\mega \hertz}$, whereas our fitted factors give $\delta \nu_\text{SMS} ^{88,86} = \SI{-267(45)}{\mega \hertz}$ and $\delta \nu_\text{FS} ^{88,86} = \SI{2(42)}{\mega \hertz}$. This shows only a rough agreement, although this level of agreement is not unexpected when compared with similar investigations \cite{Aspect1991IsotopeShiftSr}.

\subsection{Hyperfine splitting}
\label{SubSec:HyperFineStructure}

The fermionic isotope $^{87}$Sr has a nuclear spin of $I=9/2$. Hence, the \StateOneDTwo~state possesses a hyperfine structure with five levels with total angular momentum $F = \{ 5/2, 7/2, 9/2, 11/2, 13/2 \}$, whose splitting was measured to the $\si{\kilo \hertz}$ level by radio frequency spectroscopy \cite{Grundevik1983}. Accordingly, the \StateEightPOnePOne~state exhibits a hyperfine structure of three levels with $F = \{ 7/2, 9/2, 11/2 \}$. We thus expect to observe at most nine allowed transitions, similar to the experimental study of the \RepumpTransitionSixP~transition in Ref.~\cite{Grundevik1983}. However, in our measurements we observe only four clearly separated resonance peaks in a narrow range of about $\SI{150}{MHz}$, see Fig.~\ref{Fig:Spectroscopy}. In order to ensure that we do not omit other resonances, we scan a frequency range around the four peaks from $-7$ to +$\SI{+3}{GHz}$. This is large compared with the hyperfine splitting of the strongly perturbed \StateSixPOnePOne~state, which spans about $\SI{1.5}{GHz}$ \cite{Grundevik1983}, so the splitting of the \StateEightPOnePOne~state should be encompassed by our scan. We conclude that the hyperfine splitting of the \StateEightPOnePOne~state is much smaller than for the \StateSixPOnePOne~state, and instead much closer to the one of the \StateFivePOnePOne~state \cite{Xu2003}. This is compatible with the presence of four separated transitions close to each other, meaning that we do not probe only a single hyperfine state of \StateEightPOnePOne, which would otherwise give at most three allowed transitions. We therefore suppose that the nine expected transitions are located within the narrow $\SI{150}{MHz}$ range shown in Fig.~\ref{Fig:Spectroscopy} and that we cannot resolve the individual peaks due to line broadenings. 

However, we can determine boundaries for the magnetic dipole and electric quadrupole coupling coefficients A and B in the Casimir formula \cite{Casimir1936, Mortensen2004Isotope} 
\begin{equation}
    \Delta E_\text{F} = \Delta \nu + \frac{\text{A}}{2} \text{C} + \frac{\text{B}}{4} \frac{3/2\text{C}(\text{C}+1)-2\text{I}(\text{I}+1)\text{J}(\text{J}+1)}{(2\text{I}-1)(2\text{J}-1)\text{I}\text{J}},
\end{equation}
where $\text{C}=\text{F}(\text{F}+1)-\text{I}(\text{I}+1)-\text{J}(\text{J}+1)$ and $\Delta \nu$ is the isotope shift for \Sr{87}. Knowing that the splitting in the \StateOneDTwo~manifold between the lowest and highest state, with $\text{F}=13/2$ and $\text{F}=5/2$ respectively, is about $\Delta \nu_\text{splitting}^{1D2} = \SI{120}{MHz}$ \cite{Grundevik1983} and considering that we measure a signal for \Sr{87} over a range of about $\SI{150}{MHz}$, we can conclude that the energies of the three hyperfine states of \StateEightPOnePOne~are located within a maximum spread of $\sim \SI{30}{MHz}$. This translates into A and B coefficients within $A \in [-3,3] \, \si{\mega \hertz}$ and $B \in [-30,30] \, \si{\mega \hertz}$. Let's note that, owing to the detuning of the 2D MOT beams, the \StateFivePOnePOne~hyperfine states $\text{F}=11/2$ and $\text{F}=9/2$ are likely to be more strongly addressed than $\text{F}=7/2$, which is off resonant by about two times the \BlueMotTransition~transition linewidth. This means that the population in the state $\text{F}=5/2$ of \StateOneDTwo~is likely very small. In this case, the maximum spread for the three hyperfine states of \StateEightPOnePOne~is $\sim \SI{70}{MHz}$, and we have $A \in [-8,8] \, \si{\mega \hertz}$ and $B \in [-70,70] \, \si{\mega \hertz}$.

Having laid these conservative boundaries, we now attempt to find fitting values of A and B that explain the distribution of the measured transitions. We take into account all nine possible transitions and assume that the four peaks observed are composed of one or more of these transitions, and that all transitions contribute to the spectrum. For lack of better insight, we consider all possible permutations of these nine transitions to the four lines, and fit for each permutation a set of A, B and $\Delta \nu$. Each set obtained comes with an evaluation of the reduced chi square $\Tilde{\chi}^2$, which gives an estimation of the goodness of the fit. As expected for our spectrum with only a few peaks and not all resolved, none of the $10^4-10^5$ permutations clearly stands out. We thus select the few tens of models with low values of $\Tilde{\chi}^2$ and calculate from these sets mean values and standard deviations for A and B. When we assume the presence of atoms in the $\text{F}=5/2$, \StateOneDTwo~state, we obtain $A = \SI{-6(4)}{MHz}$ and $B = \SI{10(30)}{MHz}$. If, as justified previously, we neglect the population in $\text{F}=5/2$, we obtain significantly lower best $\Tilde{\chi}^2$ values (although still significantly higher than 1) and the coefficients $A = \SI{-4(5)}{MHz}$ and $B = \SI{5(35)}{MHz}$, which should thus be considered the more reliable set of results.

Finally, the small frequency spread of the \RepumpTransition~lines is advantageous for using this transition for repumping. Other repumping schemes for fermionic strontium with spread of hundreds of $\si{\mega \hertz}$ can be technologically demanding if one wants to address all potentially populated $\text{F}$ states \cite{Stellmer2014ReservSpectro}, e.g. by using several acousto-optic modulators or a chirp over a large mode-hop free frequency range. The small spread for our scheme should greatly simplify the task and provide efficient repumping.\\

\section{Repumping}
\label{Sec:Repumping}

Now that we have spectroscopically determined the properties of the \RepumpTransition~transition, we address it with our $\SI{448}{nm}$ laser in order to repump atoms in the 2D MOT that decayed into the \StateOneDTwo~state, and thus tackle this loss mechanism.

To quantify the efficiency of the repumping process, we do not rely on fluorescence imaging of the 2D MOT in the upper chamber that we previously used. Instead, we use absorption imaging of the \RedMotTransition~3D MOT occupying the lower chamber, as described in Sec.~\ref{Sec:ExperimentalProtocol}. Measuring the atom number evolution inside the 3D MOT while applying the repumper on the 2D MOT cloud allows us to observe the evolution of the flux of atoms captured by the 3D MOT. This flux is directly proportional to the flux of atoms exiting the 2D MOT, which we verified experimentally. The repumper beam is still aligned to the center of the 2D MOT and the laser beam intensity of $16.5$ times the transition's saturation intensity produces a power broadening that helps addressing the atoms despite their Doppler and Zeeman shifts. We verify that this power is sufficient to reach the saturation of the repumping effect. Let us also note that, by using a different orientation of the repumping beam, angled at $\SI{45}{\degree}$ from the Zeeman slower beam and shining only on the 2D MOT, we have verified independently that the observed gain in atom flux originates solely from a repumping effect at the MOT location, and not within the Zeeman slower region.

\begin{figure}[tb]
	\centering
	\includegraphics[width=0.45\textwidth]{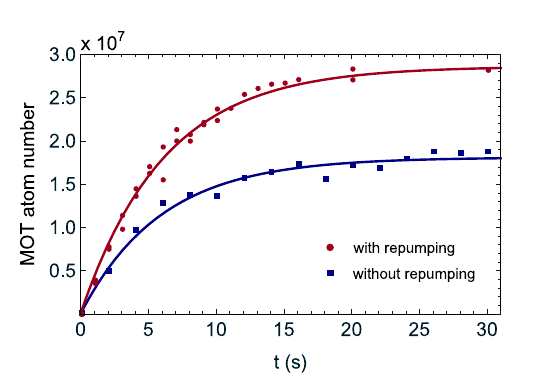}
	\caption{Evolution of the atom number in the 3D MOT as a function of the loading time, with and without the $\SI{448}{nm}$ repumping laser addressing the \RepumpTransition~transition. The $\SI{448}{nm}$ laser shines onto the 2D MOT of the upper chamber, see Fig.~\ref{Fig:setup}. We observe an increase in atom number, which comes from a $\SI{63}{\percent}$ increase in the flux of atoms leaving the 2D MOT. Lines are fit to the data using Eq.~\ref{eq:expo_fit}.} 
	\label{Fig:blueMOTrepumper}
\end{figure}

We quantify the effect of the repump laser by switching on all lasers needed for cooling and trapping the \Sr{88} isotope in the 2D and the 3D MOTs, and waiting for a varying loading time before taking an absorption image of atoms in the 3D MOT. We perform this measurement twice: once without and once with the $\SI{448}{nm}$ repump laser shining onto the 2D MOT region. For these measurements, we make sure to reduce the flux of atoms entering the 2D MOT, by adjusting the oven temperature and the parameters of the previous laser cooling stages. This is necessary to avoid additional loss mechanisms, e.g. the spilling of atoms outside of the laser beams or a too (optically) dense sample leading to a smaller restoring force or light-assisted inelastic collisions, both in the 2D and the 3D MOTs. We obtain the loading curves of Fig.~\ref{Fig:blueMOTrepumper}, which we fit using the function
\begin{equation}
    N_\text{MOT}(t) = \frac{\Phi}{\Gamma} \cdot \left( 1-e^{-\Gamma t} \right),
    \label{eq:expo_fit}
\end{equation}
where $N_\text{MOT}(t)$ is the time-dependent atom number in the 3D MOT, $\Phi$ is the flux of captured atoms and $\Gamma$ is the one-body loss rate. We obtain the fit parameters given in Table~\ref{Tab:MOTLoadingParams}. We see that with the addition of the repumping laser, the flux of atoms exiting the 2D MOT is increased by $\SI{63}{\percent}$.

\begin{table}[H]
	\centering
	\begin{tabular}{l|r|r}
     & $\Phi$ ($\times 10^7 \, \mathrm{atom} \, \si{ \per \second}$) & $\Gamma$  ($\si{\per \second}$)\\
	\hline
	Repumping beam OFF  & $\SI{0.30(2)}{}$  & $\SI{0.17(2)}{}$ \\
	Repumping beam ON   & $\SI{0.46(2)}{}$  & $\SI{0.16(1)}{}$ \\
	\end{tabular}
	\caption{Results of the fitted 3D MOT loading curves of Fig.~\ref{Fig:blueMOTrepumper}, with and without the repumping laser beam, which is tuned to the \RepumpTransition~transition and shining on the 2D MOT region. The error bars are the standard errors from the fits.}
	\label{Tab:MOTLoadingParams}
\end{table}

Owing to the inherently lossy nature of our 2D MOT, it is challenging to have an absolute determination of how well the losses via \RepumpTransition~are taken care of. Indeed, atoms that do not decay to the \StateOneDTwo~state are traveling fast across the 2D MOT and naturally exit it to continue toward the lower vacuum chamber. This exit can be considered a one-body loss mechanism for the 2D MOT, and it dominates over all other potential decay mechanisms. According to the known transitions between the electronic levels of strontium \cite{Sansonetti2010}, a majority of the atoms optically pumped to the \StateEightPOnePOne~state will either come back to \StateOneDTwo~or reach \StateOneSZero~in a few hundreds of nanoseconds, see Tab.~\ref{Tab:BranchingRatios}. This should allow efficient repumping of atoms back into the \BlueMotTransition~cooling cycle. However, this only takes into account transitions between singlet electronic states, while the losses toward, potentially long-lived, triplet states are unknown, see e.g. Refs.~\cite{Kurosu1990LaserCoolCaSr, Vogel1999PhDThesis, Bidel2022PhDThesis}. In order to estimate whether the effect of our repumping scheme on our apparatus is to repump atoms with a high probability or to repump them fast, thus increasing the mean restoring force within the 2D MOT, we compare our scheme to the commonly used \RepumpRedTransitions~repumping scheme at $\SI{679}{nm}$ and $\SI{707}{nm}$ \cite{Dinneen1999}. This scheme has the advantage of repumping all atoms back to \StateOneSZero~via the \StateThreePOne~state, but it requires a long wait time for atoms to decay from the \StateOneDTwo~state. Therefore, this scheme should not be particularly suited to our 2D MOT acting on fast moving atoms. Indeed, applying the 679nm and 707nm repumping scheme on our 2D MOT, we measure an increase of atoms in the 2D MOT cooling cycle of less than $\SI{40}{\percent}$, significantly lower than the $\SI{60}{\percent}$ improvement gained using our, likely lossier but very fast, $\SI{448}{\nano \meter}$ scheme. This indicates that the high repumping rate of our scheme is an important factor in our apparatus. However, one would have to use a 3D MOT on \BlueMotTransition~with slow loss rates to be able to estimate quantitatively the repumping efficiency and the amount of losses via triplet states.

\section{Conclusion}
\label{Sec:Conclusion}
In conclusion, we proposed and experimentally demonstrated a fast and efficient repumping scheme for improved laser cooling and imaging of strontium atoms on the \BlueMotTransition~transition. With our scheme, we can recycle atoms that spontaneously decay to the \StateOneDTwo~state directly back into the cooling cycle, by addressing the \RepumpTransition~transition. The demonstrated scheme stands out due to its very fast repumping cycle, having a characteristic decay time back into the main cooling cycle shorter than $\SI{100}{\nano \second}$ for $\gtrsim \SI{96}{\percent}$ of the atoms. We implemented this scheme with a single, low-cost, external-cavity diode laser at $\SI{448}{nm}$ and could increase the atom number in our 2D MOT by $\SI{60}{\percent}$ compared to the case without repumping. Furthermore, we performed spectroscopy on the \RepumpTransition~transition and measured its absolute frequency $\nu_{\mathrm{^{88}Sr}} = (668917515.3 \pm 4.0 \pm 25) \, \si{MHz}$. Moreover, we determined the isotope shifts between the four stable isotopes of strontium and infer the specific mass and field shift factors $\delta \nu_\text{SMS} ^{88,86} = \SI{-267(45)}{\mega \hertz}$ and $\delta \nu_\text{FS} ^{88,86} = \SI{2(42)}{\mega \hertz}$, respectively. Finally, from our measurements with the fermionic isotope $^{87}$Sr, we obtain the magnetic dipole coupling coefficient $A = \SI{-4(5)}{MHz}$ and the electric quadrupole coupling coefficient $B = \SI{5(35)}{MHz}$ for the hyperfine structure of the \StateEightPOnePOne~state.

In the future, our scheme might help to improve the performances of cold-strontium-based quantum devices such as clocks, atom interferometers, and optical tweezer arrays thanks to better laser cooling and imaging on the \BlueMotTransition~transition. It could also be helpful for the implementation of shelving schemes, where atoms are stored in the \StateThreePZeroTwo~states, while the other \StateOneSZero~part of the qubit state is interrogated with the \BlueMotTransition~transition \cite{Okuno2022TweezerYb1S03P2Qbit, Pagano2022TweezerRydbergSr, Trautmann2023SpectroSr1S03P2}. Thanks to the fast removal of atoms that have fallen into the \StateOneDTwo~state, this repumping scheme would avoid contamination of the qubit state.

\begin{acknowledgments}
\medskip \noindent
We thank the team of Rene Gerritsma for providing access to the wavemeter and helping with its calibration. We thank Vincent Barbé for helping with the wavemeter calibration. We thank Alex Urech for insightful discussions on optical tweezer arrays. We thank Junyu He for providing some technical information about the experimental setup. We thank the NWO for funding through Vici grant No. 680-47-619 and the European Research Council (ERC) for funding under Project No. 615117 QuantStro. This project has received funding from the European Union’s Horizon 2020 research and innovation programme under grant agreement No 820404 (iqClock project). B.P. thanks the NWO for funding through Veni grant No. 680-47-438 and C.-C.C. thanks support from the MOE Technologies Incubation Scholarship from the Taiwan Ministry of Education. 

\noindent
\textbf{Author contribution} J.S, S.B., B.P. set up the laser system. J.S., C.-C.C., and R.G.E. conducted the experiments and the data collection. J.S. and B.P. analysed the data. B.P., S.B. and F.S. supervised the project. F.S. acquired funding. J.S. and B.P. wrote the manuscript.

\noindent
\textbf{Correspondence and requests for materials} should be addressed to F.S. Raw data and analysis materials used in this research can be found at \cite{FigshareDataPackage}.
\end{acknowledgments}

%\bibliography{biblio}


\end{document}